\documentclass[twocolumn,english,showkeys,reprint]{revtex4}
\usepackage[T1]{fontenc}
\usepackage[latin9]{inputenc}
\usepackage{textcomp}
\usepackage{amsmath}
\usepackage{amssymb}
\usepackage{graphicx}

\makeatletter

\newcommand{\lyxmathsym}[1]{\ifmmode\begingroup\def\b@ld{bold}
  \text{\ifx\math@version\b@ld\bfseries\fi#1}\endgroup\else#1\fi}

\@ifundefined{textcolor}{}
{%
 \definecolor{BLACK}{gray}{0}
 \definecolor{WHITE}{gray}{1}
 \definecolor{RED}{rgb}{1,0,0}
 \definecolor{GREEN}{rgb}{0,1,0}
 \definecolor{BLUE}{rgb}{0,0,1}
 \definecolor{CYAN}{cmyk}{1,0,0,0}
 \definecolor{MAGENTA}{cmyk}{0,1,0,0}
 \definecolor{YELLOW}{cmyk}{0,0,1,0}
 }

\@ifundefined{definecolor}
 {\usepackage{color}}{}

\@ifundefined{textcolor}{}{%
 \definecolor{BLACK}{gray}{0}
 \definecolor{WHITE}{gray}{1}
 \definecolor{RED}{rgb}{1,0,0}
 \definecolor{GREEN}{rgb}{0,1,0}
 \definecolor{BLUE}{rgb}{0,0,1}
 \definecolor{CYAN}{cmyk}{1,0,0,0}
 \definecolor{MAGENTA}{cmyk}{0,1,0,0}
 \definecolor{YELLOW}{cmyk}{0,0,1,0}
 }

\usepackage{babel}
\usepackage{babel}

\makeatother

\usepackage{babel}
\begin{document}

\title{Magnetic and structural properties of Co$_{2}$FeAl thin films grown
on Si substrate }

\author{M. Belmeguenai$^{1}$ \footnote{Email: belmeguenai.mohamed@univ-paris13.fr} , H. Tuzcuoglu$^{1}$ , M. S. Gabor$^{2}$ \footnote{Email:
mihai.gabor@phys.utcluj.ro}, T. Petrisor jr$^{2}$ , C.
Tuisan$^{2,3}$ , D. Berling$^{4}$, F. Zighem$^{1}$ and S. M.
Chérif$^{1}$}

\affiliation{$^{1}$ LSPM-CNRS, Université Paris 13, 99 avenue Jean-Baptiste Clément
93430 Villetaneuse, France }

\affiliation{$^{2}$ Center for Superconductivity, Spintronics and Surface Science,
Technical University of Cluj-Napoca, Str. Memorandumului No. 28 RO-400114
Cluj-Napoca, Romania}

\affiliation{$^{3}$ Institut Jean Lamour, CNRS, Université de Nancy, BP 70239,
F\textendash{} 54506 Vandoeuvre, France }

\affiliation{$^{4}$ IS2M (CNRS-LRC 7228), 15 rue Jean Starcky, Université de
Haute-Alsace, BP 2488, 68057 Mulhouse-Cedex, France}
\begin{abstract}
The correlation between magnetic and structural properties of Co$_{2}$FeAl
(CFA) thin films of different thickness (10 nm$<d<$100 nm) grown
at room temperature on MgO-buffered Si/SiO2 substrates and annealed
at 600$\lyxmathsym{\textdegree}C$ has been studied. XRD measurements
revealed an (011) out-of-plane texture growth of the films. The deduced
lattice parameter increases with the film thickness. Moreover, pole
figures showed no in-plane preferential growth orientation. The magneto-optical
Kerr effect hysteresis loops showed the presence of a weak in-plane
uniaxial anisotropy with a random easy axis direction. The coercive
field measured with an applied field along the easy axis direction
and the uniaxial anisotropy field increase linearly with the inverse
of the CFA thickness. The microstrip line ferromagnetic resonance
measurements for in-plane and perpendicular applied magnetic fields
revealed that the effective magnetization and the uniaxial in-palne
anisotropy field follow a linear variation versus the inverse CFA
thickness. This allows deriving a perpendicular surface anisotropy
coefficient of -1.86 erg/cm$^{2}$.

\end{abstract}


\maketitle

\section{Introduction}

Heuslers $X_{2}YZ$ ($X$ being a transition metal element, $Y$ being
another transition metal element and $Z$ being a group $III$, $IV$,
or $V$ element) are interesting group of materials due to their potential
use as one magnetic electrode in giant and tunnel magnetoresistance
devices used in magnetic memories (MRAM) {[}3{]}, in low field magnetic
sensors {[}4{]} and in microwave components for spintronic applications
{[}5{]}. The most prominent representatives of the this kind of spintronic
materials are Co-based full Heusler alloys due to their predicted
half metallic character (magnetic materials that exhibit a 100\% spin
polarization at the Fermi level) even at room temperature and due
to their high Curie temperature {[}6{]}. These make them ideal sources
of high spin polarized currents to realize high giant magnetoresistance
values. Co2FeAl (CFA) is one of the Co-based Heusler alloys having
a very high Curie temperature (1000 K) and, theoretically predicted,
a half-metallic character of their spin-split band structure. It can
provide giant tunnelling magnetoresistance (360\% at room temperature)
{[}7{]} when used as an electrode in magnetic tunnel junctions, which
makes CFA promising for practical applications. However, the crystalline
structure and the chemical order of such materials strongly influence
their magnetic and structural properties. Moreover, the substrate
material as well as the orientation of substrate and the film thickness
have an impact on the magnetic anisotropy of magnetic thin films such
as CFA because of the band hybridization and the spin-orbit interaction
at the interface. Therefore, the correlation between their magnetic
and structural properties and its dependence on film thicknesses,
for precise control of the magnetic properties required by the integration
of CFA as a magnetic electrode in spintronic devices, should be investigated.
For this purpose, microstrip ferromagnetic resonance (MS-FMR), magneto-optical
Kerr effect (MOKE) magnetometry and X-ray diffraction (XRD) techniques
are used.

\section{Sample preparation}

CFA films thin films of different thickness (10 nm$<d<$100 nm) were
grown on Si(001)/SiO2 substrates using a magnetron sputtering system
with a base pressure lower than $3\times10^{-9}$ Torr. Prior to the
deposition of the CFA films, a 4 nm thick MgO buffer layer was grown
at room temperature (RT) by rf sputtering from a MgO polycrystalline
target under an Argon pressure of 15 mTorr. Next, CFA thin films of
different thickness, were deposited at room temperature by dc sputtering
under an Argon pressure of 1 mTorr, at a rate of 0.1 nm/s. Finally,
the CFA films were capped with a a MgO(4 nm)/Ta(4 nm) bilayer. After
the growth of the stack, the structures were ex-situ annealed at 600$\lyxmathsym{\textdegree}C$
during 15 minutes in vacuum. The structural properties of the samples
have been characterized by XRD using a four-circle diffractometer.
Their magnetic static and dynamic properties have been studied by
magneto-optical Kerr effect magnetometer (MOKE) and microstrip ferromagnetic
resonance (MS-FMR) {[}8{]}, respectively.

\begin{figure*}
\includegraphics[bb=20bp 400bp 700bp 595bp,clip,width=17cm]{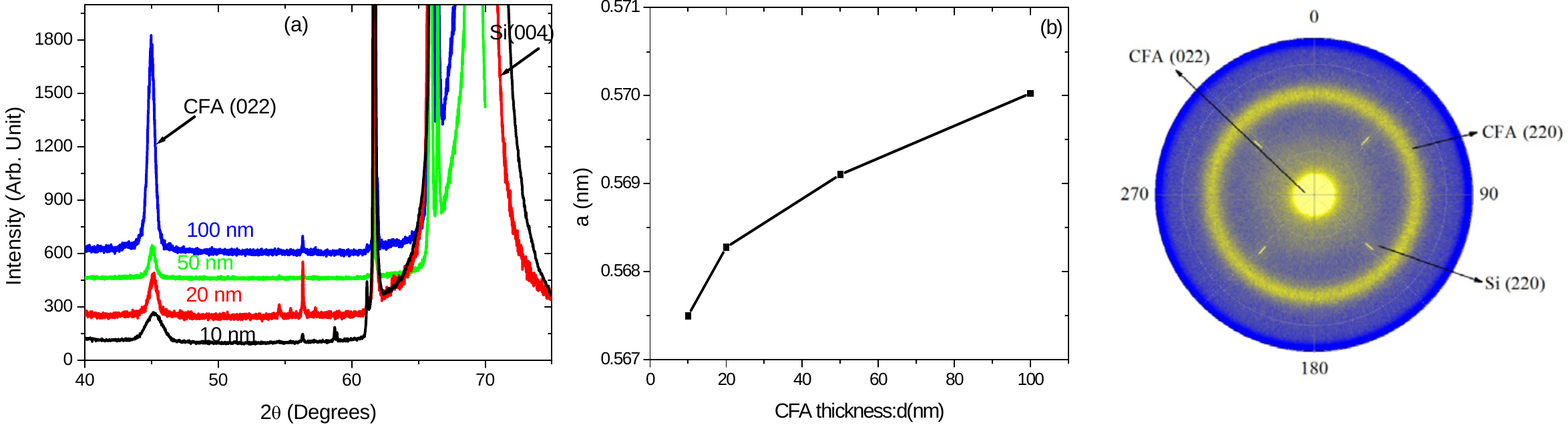}

\caption{Thickness dependence of (a) $\theta-2\theta$ patterns (b) lattice
parameter; (c) pole figure around a (022) peak of 50 nm thick CFA
film}
\end{figure*}

\section{Structural properties }

The X-rays $\theta-2\theta$ diffraction patterns for CFA thin films
of different thicknesses revealed one peak which is attributed to
the (022) diffraction line of CFA (Fig. 1a). The lattice parameter
($a$), shown on Figure 1b, increases with the increasing CFA thickness,
similarly to samples grown on MgO substrates where a direct correlation
exists with the enhancement of the chemical order {[}9{]}, however
the lattice parameters remain smaller than the reported one in the
bulk compound with the L2$_{1}$ structure (0.574 nm). Due to the
overlap of the potential (002) CFA peak with the substrate reflections
it is difficult to accurately evaluate de degree of chemical order
in our films. Pole figures around the (022) type CFA peaks (Fig. 1c)
indicates that the CFA films show a strong (011) fiber- texture with
no in-plane preferential growth direction.

\section{Magnetic properties }

\begin{figure*}
\includegraphics[bb=20bp 450bp 520bp 595bp,clip,width=17cm]{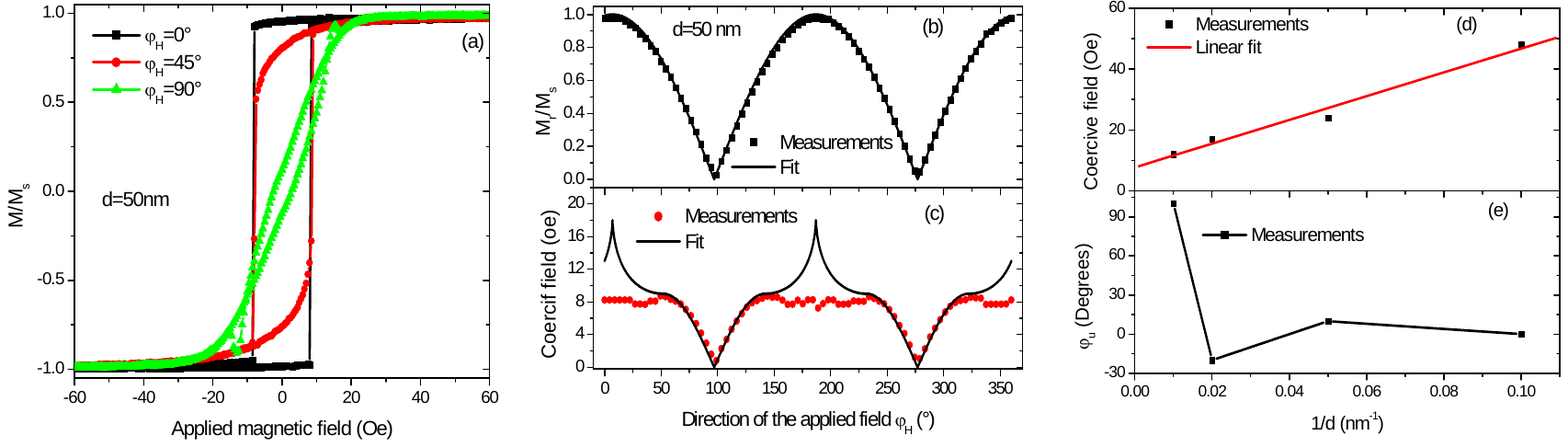}

\caption{(a) MOKE hysteresis loops and the corresponding angular of the (b)
normalized remanent magnetization (c) the coercive field of the 50
nm thick Co2FeAl thin film. (d) and (e) thickness dependence of the
easy coercive field and the magnetization easy axis respectively of
the Co2FeAl thin films.}
\end{figure*}

To analyze the experimental results, the total magnetic energy given
by equation (1) is considered.

\begin{widetext}

\begin{multline}
E=-M_{S}H\left(\cos\left(\varphi_{M}-\varphi_{H}\right)\sin\theta_{M}\sin\theta_{H}+\cos\theta_{M}\cos\theta_{H}\right)-\left(2\pi M_{S}^{2}-K_{\perp}\right)\sin^{2}\theta_{M}-\frac{1}{2}(1+\cos(2(\varphi_{M}-\varphi_{u}))K_{u}
\end{multline}

\end{widetext}

In the above expression $\varphi_{M}$ (resp. $\varphi_{H}$) represents
the in-plane (referring to the substrate edges) angle defining the
direction of the magnetization $M_{S}$ (resp. the external applied
field $H$), while $\theta_{M}$ (resp. $\theta_{H}$) is the out-of-plane
angle between the magnetization (resp. the external field $H$) and
the normal to the sample plane. $\varphi_{u}$ defines the angles
of an easy uniaxial planar axis with respect to this substrate edge.
$K_{\mathit{u}}$, and $K_{\perp}$ are in-plane uniaxial and out-of-plane
uniaxial anisotropy constants, respectively.

For an in-plane applied magnetic field, the studied model provides
the following expression (2) for the frequencies of the experimentally
observable magnetic modes:

\begin{widetext}
\begin{multline}
F_{n}^{2}=\left(\frac{\gamma}{2\pi}\right)^{2}\left(H\cos\left(\varphi_{M}-\varphi_{H}\right)+H_{u}\cos\left(\varphi_{M}-\varphi_{u}\right)+\frac{2A_{ex}}{M_{s}}\left(\frac{n\pi}{d}\right)^{2}\right)\\
\left(H\cos\left(\varphi_{M}-\varphi_{H}\right)+\frac{H_{u}}{2}\left(1+\cos\left(\varphi_{M}-\varphi_{u}\right)\right)+\frac{2A_{ex}}{M_{s}}\left(\frac{n\pi}{d}\right)^{2}+4\pi M_{eff}\right)
\end{multline}

\end{widetext}

where $\frac{\gamma}{2\pi}=g\times1.397\times10^{6}$ Hz/Oe is the
gyromagnetic factor. We introduce the effective magnetization $M_{eff}=H_{eff}/4\pi$
obtained by:
\begin{equation}
4\pi M_{eff}=H_{eff}=4\pi M_{S}-\frac{2K_{\perp}}{M_{S}}=4\pi M_{S}-H_{\perp}
\end{equation}

As experimentally observed, the effective perpendicular anisotropy
term $K_{\perp}$ (and, consequently, the effective perpendicular
anisotropy field $H_{\perp}$) is thickness dependent: $K_{\perp}$
describes an effective perpendicular anisotropy term which writes
as:
\begin{equation}
K_{\perp}=K_{\perp}^{V}+\frac{2K_{\perp}^{S}}{d}
\end{equation}

where $K_{\perp}^{S}$ refers to the perpendicular anisotropy term
of the interfacial energy density. Finally we define $H{}_{\mathit{u}}=2K_{u}/M_{S}$
as the in-plane uniaxial anisotropy field. The uniform precession
mode corresponds to n = 0. The other modes to be considered (perpendicular
standing modes: PSSW) are connected to integer values of $n$: their
frequencies depend upon the exchange stiffness constant $A_{ex}$
and upon the film thickness $d$.

In the case of an out-of-plane perpendicular applied magnetic field,
the resonance frequency is given by:
\begin{multline}
F_{\perp}=\left(\frac{\gamma}{2\pi}\right)\left(H-4\pi M_{eff}+\frac{2A_{ex}}{M_{s}}\left(\frac{n\pi}{d}\right)^{2}\right)
\end{multline}

\subsubsection{Static properties}

The overage magnetization at saturation measured by vibrating sample
magnetometer at room temperature for all samples has been found to
be $M_{s}=1000\pm50$ emu/cm$^{3}$. The typical MOKE hysteresis loops,
measured versus external magnetic field orientations with respect
to one of the substrate edges for all the samples, are represented
on figure 2a for the 50 nm thick film. The shape of the magnetization
reversal loops changes with the magnetic field orientation mainly
due to the uniaxial magnetic anisotropy. The corresponding typical
angular dependence of coercive fields ($H_{C}$) and normalized remanent
magnetizations ($M_{r}/M_{S}$) are represented on figures 2b and
2c for the 50 nm sample. The samples have a similar angular dependence
of $H_{C}$ and $M_{r}/M_{S}$ however, the $H_{C}$ magnitude and
the easy axes directions are different from each other (Fig. 2d and
2e). Apparently, the angular dependences agree with those of Stoner
Wohlfarth (coherent rotation: CR) model as shown on figures 2b and
2c. Despite of the perfect agreement for $M_{r}/M_{S}$, a significant
discrepancy for $H_{C}$ is shown around the easy axis direction between
the CR model and measurements. In fact, the coercive fields deduced
from the magnetization loops around the easy axis are smaller than
the one obtained from CR model since that they are usually determined
by domain nucleation and sample imperfection. The coercive field deducted
from hysteresis loops obtained for a magnetic field applied along
the direction of the easy axis, are shown in figure 2d as a function
of the inverse film thickness. $H_{C}$ increases linearly with the
inverse of the film thickness due maybe to the enhancement chemical
order or to the strain relaxation as the thickness increases.

\subsubsection{Dynamic properties}

\begin{figure*}
\includegraphics[bb=20bp 440bp 520bp 595bp,clip,width=17cm]{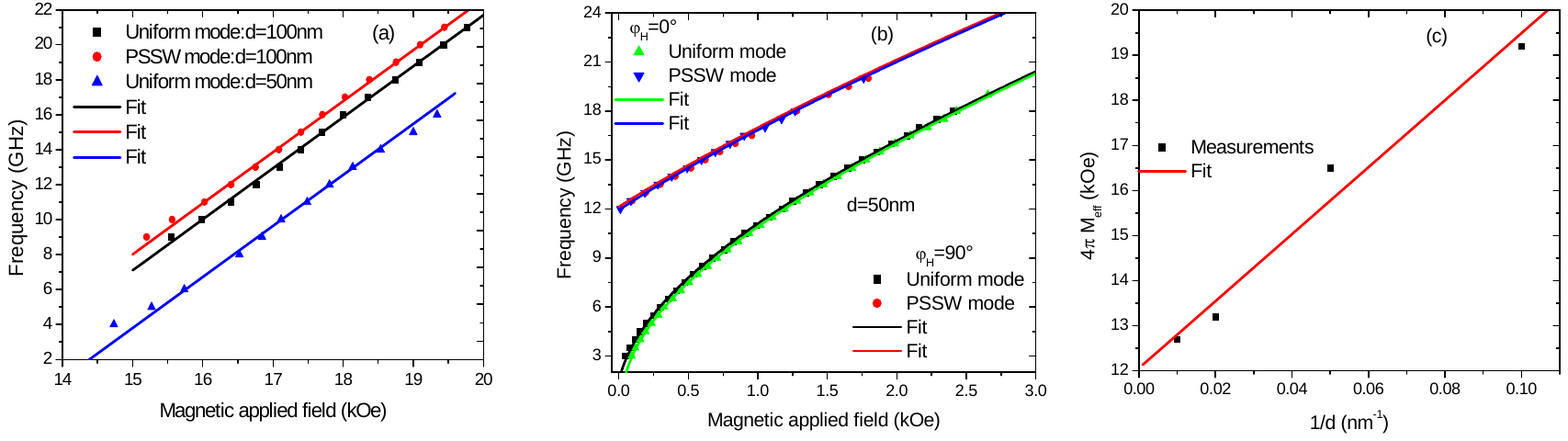}

\caption{Uniform precession and perpendicular standing spin waves (PSSW) modes
frequencies versus the (a) perpendicular and (b) in-plane applied
magnetic fields for 100 and 50 nm thick Co2FeAl thin films. The solid
lines refer to the fit suing the above mentioned model. (c) Thickness
dependence of the effective magnetization, of Co2FeAl thin films,
deduced from the MS-FMR measurements}
\end{figure*}

\begin{figure*}
\includegraphics[bb=20bp 450bp 530bp 595bp,clip,width=17cm]{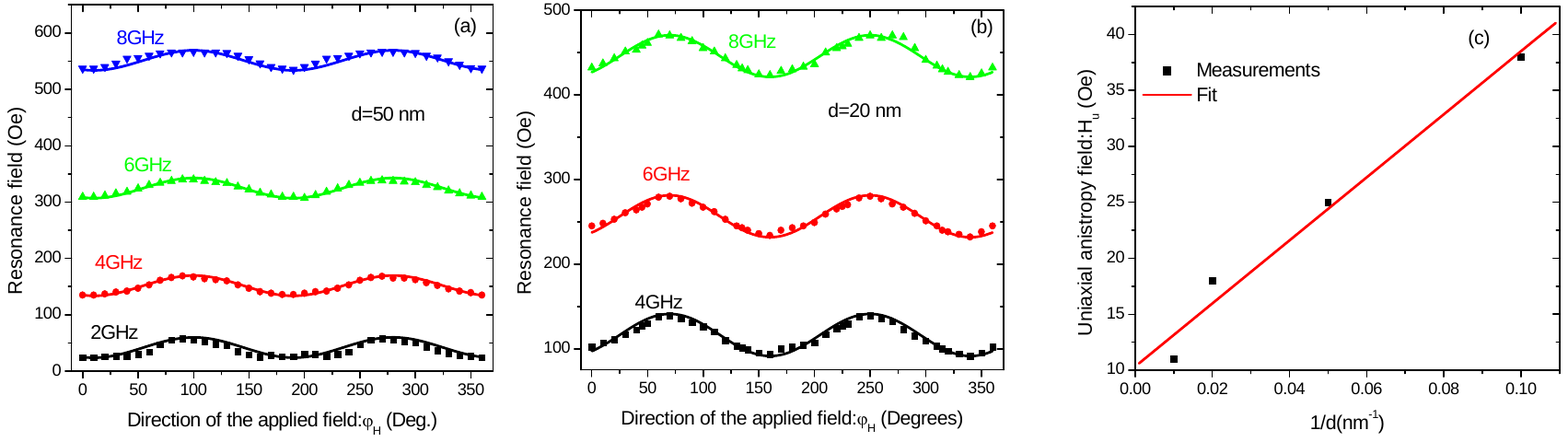}

\caption{Angular dependence of the resonance field of (a) 50 nm and (b) 20
nm thick Co$_{2}$FeAl thin films. The solid lines refer to the fit
using the above mentioned models. (c) Thickness dependence of the
uniaxial (Hu) extracted from the fit of MS-FMR measurements. The solid
lines are the linear fits.}
\end{figure*}

The MS-FMR spectra measured for in-plane and perpendicular applied
magnetic fields show the existence of a uniform mode for all samples
and a first perpendicular standing spin wave mode only for the thickest
films (100 and 50 nm thick films). Typical field dependencies of resonance
frequencies of the uniform and the first PSSW mode are shown in figures
3a and 3b respectively for magnetic fields applied in-plane and perpendicular
to the CFA film planes of various thicknesses. By fitting the data
in figure 3 to the above presented model, the gyromagnetic factor
($\gamma$), exchange stiffness constant ($A_{ex}$) and the effective
magnetization ($4\pi M_{eff}$) are extracted. The fitted $\gamma/2\pi$
and $A_{ex}$ values are found to be a constant of 29.2 GHz/T and
1.5 $\mu$erg/cm, respectively, across different samples. This exchange
constant value is in good agreement with that indicated by Trudel
et al. {[}6{]}.

Interestingly, the extracted effective magnetization from the MS-FMR
measurements, shown on figure 3c follows a linear variation versus
the inverse CFA thickness.The linear fit of the measurements allows
to determine the value of the perpendicular surface anisotropy constant:
$K_{\perp}^{S}=$-1.86 erg/cm$^{2}$. The $4\pi M_{eff}$ value when
$d$ tends to infinity, equal to 12.05 kOe, is very close to the $M_{s}$
mentioned above, taking into account experimental accuracy. This value
of the surface perpendicular anisotropy constant is in the range (1.3
erg/cm$^{2}$ {[}10{]} and 1.8 erg/cm$^{2}$ {[}11{]}), but of an
opposite sign with those reported in the MgO-based magnetic tunnel
junctions having CoFeB electrodes or in Ta/CFA/MgO multilayers {[}12{]}.
Ikeda et al. {[}10{]} attributed this perpendicular anisotropy in
the Ta/CoFeB/MgO structure to the contribution of the CoFeB/MgO interface.
Wang et al. {[}13{]} have also argued that this anisotropy is due
to the CoFeB/MgO interface based on the fact that it is only observed
when the MgO thickness overpasses 1 nm. Therefore, we conclude that
the uniaxial perpendicular anisotropy derives from surface energy
term. Its negative value favors in-plane orientation of the magnetization.
The origin of this anisotropy is most likely due to the CFA/MgO interface.
This was confirmed (note shown here) by measuring the effective magnetization
of a 10 nm CFA thick layer deposited on a SrTiO3 substrate by using
a 3 nm CoFe thick buffer layer where a $4\pi M_{eff}$ is equal to
12.5 kOe has been measured.

Figures 4a and 4b show the typical MS-FMR angular dependence of the
resonance field at different frequency for 20 and 50 nm thick films.
The MS-FMR measurements show that the angular behavior of resonance
fields is governed by the term of uniaxial anisotropy. The uniaxial
anisotropy field, presented on figure 4c, shows a similar behavior
as the coercive field and increases linearly with the CFA inverse
thickness. This uniaxial anisotropy maybe due to the substrate vicinal
structure induced by the miscut, due to the enhancement of the CFA
chemical order {[}14{]} or due to the growth-induced strain. However,
the easy direction of the uniaxial anisotropy, determined with a 90$\lyxmathsym{\textdegree}$
precision since it is referenced with a respect to substrate edges,
changes with the thickness (Fig. 2e) and complicates the identification
of its origin. Therefore, a completely satisfactory interpretation
of the presence of Hu and its thickness dependency is still missing.
Assuming that the dependence of the thickness is due to uniaxial surface
anisotropy ($H{}_{\mathit{u}}=H_{u}^{V}+4K_{u}^{S}/M_{S}$), the uniaxial
surface anisotropy constant estimated from the linear fit of uniaxial
anisotropy field is $K_{u}^{S}=7\times10^{-3}$ erg/cm$^{2}$. For
CFA films grown on Si substrates, the absence of the in-plane fourfold
anisotropy, observed in the case of CFA thin films deposited on MgO
{[}9,14{]}, is in good agreement with the above-mentioned in-plane
random orientation crystal growth of the CFA films grown on Si substrates.
For all the samples, the obtained values of the magnetic parameters
corresponding to the best fits of the MS-FMR measurements have been
used to fit the MOKE measurements where a good agreement has been
obtained.

\section{Conclusion}

Co$_{2}$FeAl films of of various thicknesses (10 nm$<d<$100 nm)
were prepared by sputtering on a Si(001) substrates and annealed at
600$\lyxmathsym{\textdegree}C$. They show an out-of-plane epitaxial
growth with an in-plane polycrystalline structure. The out-plane lattice
parameter increases with increasing CFA due to the enhancement the
chemical order or the residual strain. MOKE hysteresis loops obtained
with different field orientations revealed that the easy axis coercive
field varies linearly with the inverse CFA thickness. The microstrip
ferromagnetic resonance has been used to study the dynamic properties.
The angular and the field dependences of the resonance field and frequency,
respectively, have been analyzed through a model based on a magnetic
energy density which, in addition to Zeeman, demagnetizing and exchange
terms, is characterized by a uniaxial anisotropy to determine the
most relevant parameters. The in-plane uniaxial anisotropy field,
present in all the samples, increases linearly with the inverse CFA
thickness. However, the uniaxial anisotropy easy axis direction changes
with the thickness. The effective magnetization shows drastically
linear increase with the inverse CFA thickness due to the enhancement
of the CFA/MgO interface quality.
\begin{acknowledgments}
This work was partially supported by POS CCE Project ID.574, code
SMIS-CSNR 12467 and CNCSIS UEFISCSU Project No PNII IDEI 4/2010 code
ID-106. \end{acknowledgments}

\end{document}